\documentclass[12pt, a4paper]{article}
\usepackage[utf8]{inputenc}
\usepackage[T1]{fontenc}
\usepackage{geometry}
\usepackage{amsmath, amssymb}
\usepackage{graphicx}
\usepackage{booktabs}
\usepackage{natbib}
\setcitestyle{authoryear,round} 
\usepackage{hyperref}
\hypersetup{colorlinks=true, linkcolor=blue, citecolor=blue, urlcolor=blue}

\title{``Don't Fall Behind'': A Unified Framework of Dynastic Survival, Two-Stage Belief Error, and the Modern Involution Trap}
\author{Dong Yang}
\date{\today}

\begin{document}

\maketitle

\begin{abstract}
We set out to solve a dual puzzle regarding reproductive strategies:

\textbf{The ``Ancient vs. Modern'' Puzzle:} Why did pre-modern elites, facing high uncertainty ($\sigma$), adopt a ``Survival'' strategy (Quantity), while modern elites adopt an ``Anxiety'' strategy (Quality)?

\textbf{The ``Class Divide'' Puzzle:} Why does modern ``involution'' manifest as a U-shaped fertility pattern, where the ``Super Rich'' and the ``Poor'' maintain fertility, while the ``Aspirational Middle Class'' faces reproductive stagnation?

We develop a unified computational framework (DP + Monte Carlo) that introduces \textit{Cognitive Heterogeneity} across classes. Our \textbf{Hybrid Model (M-H)} posits that the poor act as ``Rational Survivors'' (M1 utility, Reality parameters), while the middle/rich act as ``Biased Strivers'' (M4b utility, Belief parameters).

Our simulations yield three core findings:
\begin{enumerate}
    \item \textbf{(Finding 1) Objective Rationality:} We confirm that the ``Survival'' (M1) strategy is objectively rational whenever risk exceeds a low threshold ($\sigma > 0.45$). Given that real-world risk is massive ($\sigma_{Real} \approx 4.9$), the modern ``Quality'' strategy is objectively fragile.
    \item \textbf{(Finding 2) The Behavioral Mechanism (The Strivers' Trap):} The trap for the Middle/Rich ($B \ge 200$) is driven by a ``Two-Stage Belief Error.'' They are first ``baited'' by a Causal Error (underestimating risk, $\sigma_{Belief} \approx 0.4$) to enter the status game, and then ``trapped'' by a Marginal Error (underestimating returns, $\alpha_{Belief} \approx 0.5$) which triggers a stop in fertility.
    \item \textbf{(Finding 3) The Cognitive Divide and the ``Competence Trap'':} Most critically, the U-shape is driven by the cognitive divide. The Poor ($B < 200$) escape the trap by retaining a ``Rational Survival'' strategy (M1) in the face of real high risk. Conversely, the Aspirational Middle Class ($HC \approx 12$, $B \ge 200$) is uniquely trapped by their Biased Beliefs (M4b). Their high competence raises their dynastic reference point ($R$) to a level where, under perceived low returns, restricting fertility to $N=1$ becomes the only rational choice within their biased belief system.
\end{enumerate}

\end{abstract}

\section{Introduction}

Across the developed world, and particularly in East Asia, a profound demographic shift is reshaping the social fabric: the emergence of an ``Involution Trap'' characterized by intense parental anxiety, spiraling educational investments, and a sharp decline in fertility. Empirical studies confirm the pervasiveness of this anxiety, driven by competitive educational environments and the perceived high stakes of social mobility \citep{ZhuLuo2023}. This phenomenon presents a dual puzzle for economic historians and demographers alike.

\textbf{The First Puzzle: ``Ancient vs. Modern'' Strategy.} Historically, elites faced with high environmental uncertainty (e.g., pre-modern emperors) adopted a ``Survival Strategy'' (M1): gambling on quantity (High-N, Low-I) to maximize the probability that at least one heir would survive. In contrast, modern elites, despite facing their own form of high uncertainty (e.g., status volatility), have universally shifted to an ``Anxiety Strategy'' (M2/M4b): gambling on quality (Low-N, High-I). Why has the optimal response to risk flipped from ``diversification'' to ``concentration''?

\textbf{The Second Puzzle: The ``Class Divide''.} Within modern society, this ``involution'' is not uniform. Empirical observation and demographic research across various contexts suggest a paradoxical ``U-shaped'' fertility pattern \citep{Kirdar2016}. On one end, the ``Super Rich'' (High-Budget) escape the trap through abundant financial capital. On the other end, the Poor also appear to escape, maintaining higher fertility. The ``Fertility Trap'' specifically targets the ``Aspirational Middle Class''—those with high human capital (High HC) but constrained budgets. Why does ``competence'' (High HC) paradoxically lead to reproductive stagnation, while poverty seemingly leads to reproductive resilience?

\subsection{Departure from Standard Models}

To explain these puzzles, standard economic models typically rely on the ``Quality-Quantity Trade-off'' driven by opportunity costs \citep{BeckerLewis1973}. The relationship between fertility timing and returns to education is complex, with empirical evidence suggesting polarizing effects across different education levels \citep{LiuHu2018}. More recently, \citet{DoepkeZilibotti2017, DoepkeZilibotti2019} have argued that the rise of intensive parenting is a rational response to high economic inequality. In their framework, parents correctly perceive the high returns to education and rationally choose to invest heavily in fewer children.

We challenge this ``Rational Adaptation'' hypothesis and the assumption of cognitive homogeneity that underpins it.

\begin{itemize}
    \item \textbf{Cognitive Heterogeneity:} Crucially, we depart from existing models by proposing that cognitive processes—both the perception of parameters and the fundamental utility framework—are endogenous to one's socio-economic status. We argue that the modern fertility decline is not a universal rational adjustment, but a behavioral trap driven by systematic belief errors that specifically affect those who enter the ``Status Game.''
    \item \textbf{Rationality vs. Bias:} Unlike Doepke \& Zilibotti’s agents who optimize based on correct parameters, we model a society divided between ``Rational Survivors'' and ``Biased Strivers.'' The ``Strivers'' (Middle/Rich) suffer from a ``Two-Stage Belief Error'' that lures them into a strategy (M4b) that is objectively dominated in the real high-risk world. Conversely, the ``Survivors'' (Poor) retain a rational perception of reality and adopt the M1 strategy.
    \item \textbf{The Role of Aspirations:} We introduce the concept of the ``Competence Trap.'' We show that high human capital is not just a productive asset but a driver of limiting aspirations ($R$). Drawing on the literature on endogenous aspiration formation \citep{GenicotRay2017}, we model aspirations as reference points that rise with one's own status. This mechanism, when combined with biased beliefs, explains the counter-intuitive finding that ``competent'' elites are trapped.
\end{itemize}

Our model serves as a crucial ``Proof of Concept.'' We demonstrate that by incorporating psychologically plausible cognitive heterogeneity—grounded in the literature on self-serving bias, reference group effects, and the salience of existential risk \citep{Bordalo2013}—the model precisely replicates the complex, U-shaped fertility patterns observed in reality, a result that competing uniform rational or institutional models fail to generate.

\subsection{Our Theory and Findings}

We develop a unified computational framework utilizing a \textbf{Hybrid Model (M-H)} that models the dynastic decision as an interplay between Risk ($\sigma$), Anxiety ($\lambda$), Return-to-Effort ($\alpha$), and Parental Aspirations ($R$), characterized by a sharp cognitive boundary at $B=200$.

\begin{itemize}
    \item \textbf{The Poor ($B<200$):} Utilize M1 (Survival) utility and perceive Reality ($\sigma=4.9, \alpha=0.665$).
    \item \textbf{The Middle/Rich ($B \ge 200$):} Utilize M4b (Anxiety) utility and perceive Beliefs ($\sigma=0.4, \alpha=0.5$).
\end{itemize}

Our simulations yield three core findings. \textbf{Finding 1} establishes that in the high-risk real world ($\sigma_{Real} \approx 4.9$) \citep{Agostinelli2025}, the ``Survival'' strategy is objectively dominant. \textbf{Finding 2} identifies the ``Two-Stage Belief Error'' (Causal Error and Marginal Error) that traps the Strivers. \textbf{Finding 3} reveals that the U-shape is driven by the cognitive divide: the poor act rationally to survive, while the Aspirational Middle Class is trapped by high expectations ($R$) colliding with pessimistic belief parameters.

\section{The Model Framework}

Our unified framework analyzes the optimal choice of fertility ($N$) and investment per child ($I$) by modeling the dynastic decision process. Crucially, we introduce \textbf{Cognitive Heterogeneity}, positing that modern society is divided into distinct cognitive regimes based on socio-economic status.

\subsection{Production and Constraints}

All models share an identical Human Capital Production Function, representing the objective economic environment. We follow the standard specification \citep{BeckerTomes1976, Agostinelli2025}:
\begin{equation}
    HC_i = A \cdot I_i^\alpha \cdot \varepsilon_i
\end{equation}
where $HC_i$ is the child's human capital, $A$ is total factor productivity, $I_i$ is parental investment, and $\alpha$ is the elasticity of investment (return to effort). $\varepsilon_i$ represents a stochastic shock (``luck''), assumed to be log-normally distributed: $\ln(\varepsilon_i) \sim \mathcal{N}(0, \sigma^2)$.

Parents maximize utility subject to a standard Budget Constraint:
\begin{equation}
    \sum_{i=1}^N (K_i + I_i) \le B
\end{equation}
where $B$ is the total household budget, and $K_i$ is the fixed cost per child. In the dynamic implementation (M4b), we allow for gender-specific costs ($K_m, K_f$), although in our baseline calibration, we assume $K_m = K_f$.

\subsection{The Utility Regimes}

We utilize three distinct utility functions representing different strategic environments or behavioral modes.

\textbf{Model 1 (M1): The ``Survival'' (Emperor and the Poor) Model.} This static model captures the strategy under high existential risk. The goal is ``Dynastic Survival'': ensuring that at least one heir meets a critical status threshold ($R_{survival}$). Utility is binary and depends only on the maximum outcome among all children:
\begin{equation}
    U_{M1} = P(\max(HC_1, \dots, HC_N) \ge R_{survival})
\end{equation}
M1 prioritizes diversification (Quantity) to hedge against risk.

\textbf{Model 3 (M3): The ``Altruism'' (Becker) Model.} This static model serves as our rational, anxiety-free control group. Parents are purely altruistic, deriving utility from the total human capital of their children \citep{BeckerLewis1973}.
\begin{equation}
    U_{M3} = E\left[\sum_{i=1}^N \ln(HC_i)\right]
\end{equation}

\textbf{Model 4b (M4b): The ``Status Anxiety'' (Strivers) Framework.} This framework models the modern ``Involution Trap.'' It is governed by Prospect Theory \citep{KahnemanTversky1979} and consistent with evidence on relative income concerns \citep{Luttmer2005}. Parents evaluate outcomes as gains or losses relative to a specific status reference point ($R_i$), and they are loss-averse ($\lambda > 1$). The per-child utility $v(g_i)$ is defined over the gain/loss $g_i = \ln(HC_i) - \ln(R_i)$:
\begin{equation}
    v(g_i) = \begin{cases} g_i & \text{if } g_i \ge 0 \\ \lambda g_i & \text{if } g_i < 0 \end{cases}
\end{equation}
Total utility is the expected sum of per-child utilities: $U_{M4b} = E\left[\sum_{i=1}^N v(g_i)\right]$. We deploy this dynamically, incorporating gendered reference points ($R_{son} = HC_{parent}$ vs. $R_{dtr} = HC_{average}$).

\subsection{The Hybrid Model (M-H) and the Cognitive Boundary}

Our main theoretical framework posits that modern society is characterized by a \textbf{Cognitive Class Divide}. We model this using a Hybrid Model (M-H) where the utility function and the perceived parameters switch based on a critical budget threshold ($B_{crit}=200$).
\begin{equation}
    \text{Strategy}(B) = \begin{cases} \text{M1 (Rational Survivor)} & \text{if } B < 200 \\ \text{M4b (Biased Striver)} & \text{if } B \ge 200 \end{cases}
\end{equation}
The \textbf{Rational Survivors} ($B < 200$) utilize the M1 utility function and perceive ``Reality'' parameters. The \textbf{Biased Strivers} ($B \ge 200$) utilize the M4b framework and operate under ``Belief'' parameters.

\subsection{The Dynamic Programming Structure (for M4b Agents)}

The M4b model formalizes the fertility choice for the \textbf{Biased Strivers} ($B \ge 200$) as a sequential optimization problem solved via Dynamic Programming (DP) with a maximum fertility constraint ($N_{max}=3$).
The decision process unfolds in stages defined by the current number of children $N$.
At each stage, the state $S$ is defined by the composition of the existing family: $S = (N_m, N_f)$, where $N_m$ and $N_f$ are the number of sons and daughters, respectively.

The Bellman equation defines the recursive relationship:
\begin{equation}
    V_N(S) = \max \{ V^{Stop}_N(S), V^{Grow}_N(S) \}
\end{equation}
Parents choose the action (Stop or Grow) that maximizes their value. We solve the model using backward induction.

\begin{itemize}
    \item \textbf{The Value of Stopping ($V^{Stop}$):}
    If parents choose to stop, they optimize the investment ($I$) for their current children, subject to the remaining budget $B_{rem} = B - (N_m K_m + N_f K_f)$. We optimize $I_m$ and $I_f$ independently subject to the budget constraint:
    \begin{equation}
        V^{Stop}_N(S) = \max_{I_m, I_f} \left\{ E[U_{M4b}(N_m, N_f, I_m, I_f)] \right\} \quad \text{s.t. } (N_m I_m + N_f I_f) \le B_{rem}
    \end{equation}
    This approach ensures that parents rationally allocate more resources to the child with the higher aspiration gap (typically the son in the dynastic model), rather than being forced into a suboptimal uniform allocation. The expectation $E[\cdot]$ is calculated using high-precision Monte Carlo simulation to account for the stochastic shocks ($\varepsilon$) and loss aversion ($\lambda$).

    \item \textbf{The Value of Growing ($V^{Grow}$):}
    If parents choose to grow (only possible if $N < 3$), they anticipate the expected value of the next stage $V_{N+1}$, accounting for the 50\% probability of having a son or a daughter:
    \begin{equation}
        V^{Grow}_N(S) = 0.5 \cdot V_{N+1}(N_m+1, N_f) + 0.5 \cdot V_{N+1}(N_m, N_f+1)
    \end{equation}
\end{itemize}

\noindent \textbf{Backward Induction Process:}
\begin{itemize}
    \item \textbf{N=3 (Terminal Stage):} Since $N_{max}=3$, the only option is to stop. $V_3(S) = V^{Stop}_3(S)$.
    \item \textbf{N=2:} Parents compare $V^{Stop}_2(S)$ with the expected value of moving to N=3, $E[V_3(S')]$.
    \item \textbf{N=1:} Parents compare $V^{Stop}_1(S)$ with the expected value of moving to N=2, $E[V_2(S')]$.
\end{itemize}

\section{Model Calibration: Cognitive Heterogeneity and the Hybrid Framework}

Our central thesis posits that the modern ``U-shaped'' fertility pattern is driven by \textbf{Cognitive Heterogeneity}. We argue that the perception of the economic environment and the fundamental utility framework are not uniform across society but are endogenous to socio-economic status.

To operationalize our Hybrid Model (M-H), we introduce a \textbf{Cognitive Boundary} at $B_{crit}=200$. This threshold represents the point at which families possess sufficient resources to enter the competitive ``Status Game.''
\begin{itemize}
    \item \textbf{The Rational Survivors ($B < 200$):} The Poor utilize the M1 (Survival) utility function and operate in the ``Reality'' world ($\alpha_{Real}, \sigma_{Real}$).
    \item \textbf{The Biased Strivers ($B \ge 200$):} The Middle/Rich utilize the M4b (Anxiety) utility function and operate in the ``Belief'' world ($\alpha_{Belief}, \sigma_{Belief}$).
\end{itemize}

\subsection{The ``Reality'' Parameters (The Objective World)}

The ``Reality'' parameters describe the objective economic environment.

\textbf{Return to Effort ($\alpha_{Real}$):} We adopt the consensus estimate for the elasticity of human capital production. Following \citet{BeckerTomes1976} and \citet{CunhaHeckman2007}, and consistent with the meta-analysis by \citet{Agostinelli2025}, we set the objective return to effort at:
\begin{equation}
    \alpha_{Real} = 0.665
\end{equation}

\textbf{Uncertainty/Luck ($\sigma_{Real}$):} We utilize the high-precision estimates from \citet{Agostinelli2025}, who found that the vast majority of variance in human capital outcomes is attributable to stochastic shocks ($\varepsilon_i$). We calibrate real-world uncertainty to match their findings:
\begin{equation}
    \sigma_{Real} \approx 4.9
\end{equation}

\subsection{The Cognitive Divide: Justifying Heterogeneity}

Why do the poor perceive this high-risk reality while the middle class does not?
\begin{itemize}
    \item \textbf{The Rationality of the Poor ($B < 200$):} For the poor, the massive real-world uncertainty ($\sigma=4.9$) is highly salient (Salience of Existential Risk). Furthermore, they have not yet achieved the level of success required to trigger the biases that underestimate risk (Absence of Self-Serving Bias).
    \item \textbf{The Biases of the Strivers ($B \ge 200$):} Families crossing the $B=200$ threshold shift into the ``Status Anxiety'' framework (M4b), aligning with observations of ``intensive parenting'' \citep{Lareau2011}. Entry into this regime activates a ``Two-Stage Belief Error.''
\end{itemize}

\subsection{The ``Belief'' Parameters (The Strivers' Trap)}

The ``Strivers'' operate under a distorted perception of reality.

\textbf{1. The Causal Error ($\sigma_{Belief}$): The Illusion of Control.} Upon achieving success, agents systematically underestimate the role of luck due to the Self-Serving Bias \citep{MillerRoss1975}. This fosters an Illusion of Control \citep{Langer1975}. We calibrate this perceived risk at:
\begin{equation}
    \sigma_{Belief} = 0.4
\end{equation}
This dramatic underestimation ($\sigma_{Belief} \ll \sigma_{Real}$) ``baits'' them into the Quality game.

\textbf{2. The Marginal Error ($\alpha_{Belief}$): The Pessimism of Competition.} While overestimating control, Strivers \textit{underestimate} returns. This is driven by Reference Group Effects \citep{Luttmer2005} and intense local competition \citep{HopkinsKornienko2004}. We calibrate the perceived return at:
\begin{equation}
    \alpha_{Belief} = 0.5
\end{equation}
This underestimation ($\alpha_{Belief} < \alpha_{Real}$) acts as the ``trap.''

\subsection{Behavioral and Structural Parameters}

\begin{itemize}
    \item \textbf{Loss Aversion ($\lambda$):} Following \citet{KahnemanTversky1979}, we set $\lambda = 2.5$ for M4b agents.
    \item \textbf{Reference Points ($R$):} We assume endogenous aspirations \citep{GenicotRay2017}. $R_{son} = HC_{parent}$ (Dynastic reference) and $R_{dtr} = HC_{average}$ (Assortative mating reference, with $HC_{average} = 5.0$).
    \item \textbf{Fixed Costs ($K$):} We set $K=32.0$ \citep{Lino2017}.
\end{itemize}

\subsection{Summary of Calibrated Parameters}

Table \ref{tab:calibration} summarizes the cognitive heterogeneity embedded in our Hybrid Model.

\begin{table}[htbp]
    \centering
    \renewcommand{\arraystretch}{1.3} 
    \setlength{\tabcolsep}{10pt}      
    \caption{Calibration of the Hybrid Model (M-H)}
    \label{tab:calibration}
    \begin{tabular}{@{} l l c c @{}} 
        \toprule
        & & \textbf{Rational Survivors} & \textbf{Biased Strivers} \\
        \textbf{Param} & \textbf{Description} & \textbf{($B < 200$)} & \textbf{($B \ge 200$)} \\
        \midrule
        Model & Strategic Framework & M1 (Survival) & M4b (Anxiety) \\
        $\alpha$ & Return to Effort & 0.665 (Reality) & 0.5 (Belief) \\
        $\sigma$ & Uncertainty/Luck & 4.9 (Reality) & 0.4 (Belief) \\
        $\lambda$ & Loss Aversion & --- & 2.5 \\
        $R_{son}$ & Reference (Son) & $HC_{parent}$ & $HC_{parent}$ \\
        $R_{dtr}$ & Reference (Dtr) & $HC_{parent}$ & 5.0 ($HC_{avg}$) \\
        $K$ & Fixed Cost & 32.0 & 32.0 \\
        \bottomrule
    \end{tabular}
\end{table}
\section{Baseline Results: The Objective Rationality of Risk}

Before deploying our main Hybrid Model, we first establish an objective baseline. This section investigates: given the true economic environment (``Reality'' parameters), what is the objectively rational reproductive strategy? We assume agents are perfectly rational and correctly perceive $\alpha_{Real}=0.665$.

\subsection{The Setup: Stylized Strategies}

We compare three archetypal strategies for a baseline middle-class family ($HC=6, B=200$):
\begin{itemize}
    \item \textbf{M1 (Survival/Quantity):} The risk-hedging strategy maximizing dynastic survival probability.
    \item \textbf{M2 (Anxiety/Quality):} A static implementation of the Status Anxiety framework.
    \item \textbf{M3 (Altruism/Becker):} The standard economic benchmark \citep{BeckerLewis1973}.
\end{itemize}

\subsection{The Impact of Risk and Strategic Divergence}

\begin{figure}[htbp]
    \centering
    \includegraphics[width=0.9\textwidth]{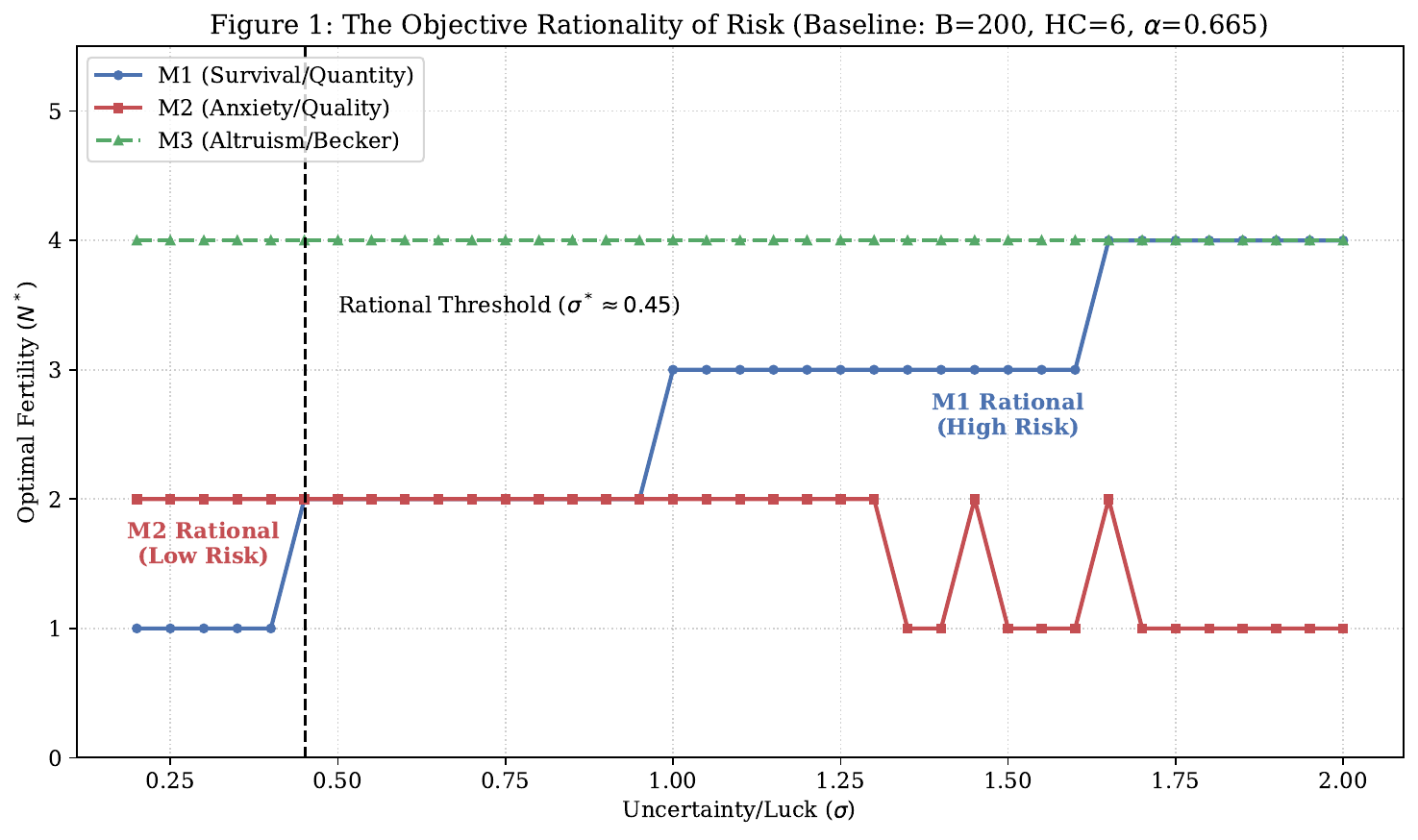}
    \caption{The Objective Rationality of Risk (Baseline: $B=200, HC=6, \alpha=0.665$). The ``Rational Threshold'' ($\sigma^* \approx 0.45$) marks where diversification becomes optimal.}
    \label{fig:fig1}
\end{figure}

Figure \ref{fig:fig1} simulates optimal fertility ($N^*$) as environmental uncertainty ($\sigma$) increases.

The simulation reveals a fundamental divergence. M1 (Survival) responds to risk with \textbf{diversification} (increasing $N$), while M2 (Anxiety) responds with \textbf{concentration} (decreasing $N$).

\textbf{The Opposing Logics of Risk Management:}
The simulation demonstrates that M1 and M2 employ fundamentally opposite strategies to manage uncertainty.
\begin{itemize}
    \item \textbf{M1 (Survival/Diversification):} The M1 strategy aims to maximize the probability that \textit{at least one} heir succeeds. When risk is low, reliance on a single child is efficient. However, as risk increases, M1 adapts by hedging against individual shocks through increased fertility (diversification).
    \item \textbf{M2 (Anxiety/Concentration):} The M2 strategy is driven by the goal of ensuring children meet the aspiration level ($R$), coupled with loss aversion ($\lambda$). As risk increases, the expected disutility from potential failures rises dramatically. To mitigate this anxiety, parents must increase investment per child ($I$), which forces a reduction in fertility (concentration).
\end{itemize}

We define the \textbf{Rational Threshold ($\sigma^*$)} as the level of uncertainty where the M1 strategy switches from $N=1$ to $N>1$. For the baseline family, $\sigma^* \approx 0.45$.

Crucially, the modern belief ($\sigma_{Belief}=0.4$) falls just below this threshold, while the real world ($\sigma_{Real} \approx 4.9$) is far above it. This confirms that in reality, the M1 strategy is objectively dominant, while the modern M2 strategy is objectively fragile.

\subsection{The Impact of Aspirations}

\begin{figure}[htbp]
    \centering
    \includegraphics[width=0.9\textwidth]{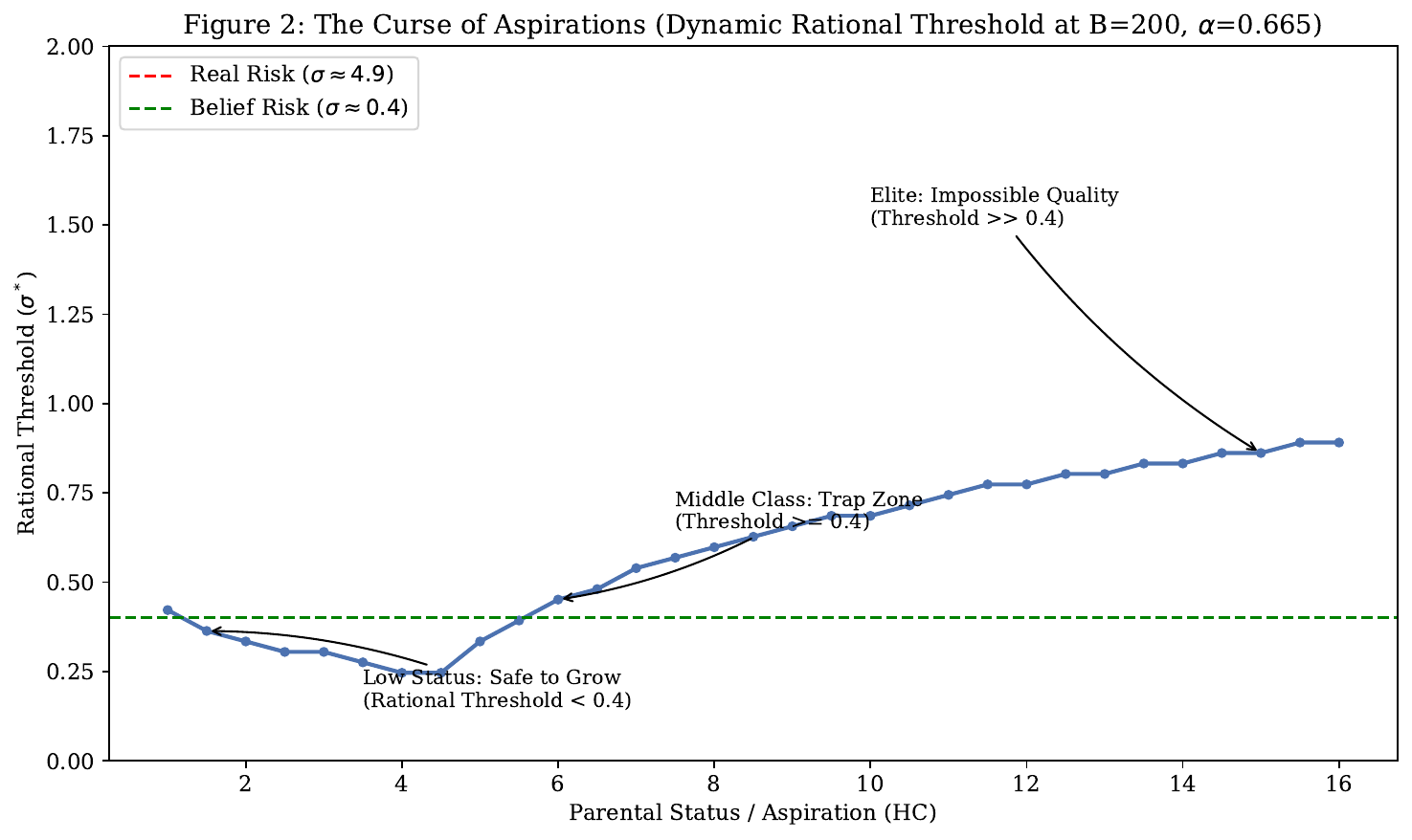}
    \caption{The Curse of Aspirations (Dynamic Rational Threshold at $B=200, \alpha=0.665$).}
    \label{fig:fig2}
\end{figure}

Figure \ref{fig:fig2} plots the Rational Threshold ($\sigma^*$) as a function of parental Human Capital.

The results reveal a non-monotonic, U-shaped relationship between parental HC and the Rational Threshold ($\sigma^*$). This pattern reflects the interplay of two opposing mechanisms that dominate at different status levels:

\begin{enumerate}
    \item \textbf{The Saturation Effect (Dominant at Low HC, 1-4):}
    The M1 utility function (Probability of Survival) is bounded at 1.0. When HC is extremely low, the aspiration goal ($R$) is trivial, and the $N=1$ strategy easily achieves a success probability near 1.0 (utility saturation). Therefore, a significant level of risk is required before the diversification benefit outweighs the cost of diluting investment. As HC rises towards 4, the risk of failure becomes salient, lowering the threshold to its minimum ($\sigma^* \approx 0.25$).
    
    \item \textbf{The Aspiration Effect (Dominant at High HC, $>4$):}
    As HC continues to rise, the aspiration goal ($R$) becomes increasingly demanding. Due to the concavity of the production function ($\alpha<1$) and the fixed budget constraint, the efficiency loss from switching from $N=1$ (concentrated investment) to $N=2$ (diluted investment) becomes very costly. This ``Aspiration Burden'' dominates the diversification motive, pushing the threshold significantly higher.
\end{enumerate}

This dynamic creates the \textbf{``Curse of Competence''} for high-status families. Nevertheless, even for the highest elites where the threshold reaches its peak, it remains far below the actual environmental risk ($\sigma^*_{Elite} \approx 0.9 \ll \sigma_{Real} \approx 4.9$).

\subsection{Implications for the Hybrid Model}

This analysis establishes that the historical shift from M1 to M4b cannot be explained as a rational adaptation to objective risk. The behavior of the Poor ($B<200$) in our Hybrid Model is objectively rational. The Middle/Rich ($B \ge 200$) must be operating under a distorted perception of reality, necessitating the ``Belief'' parameters.
\section{Main Results: The Hybrid Model and the Cognitive Divide}

In this section, we deploy our main theoretical framework—the \textbf{Hybrid Model (M-H)}—to simulate the decision-making landscape of modern parents.
\begin{itemize}
    \item \textbf{$B < 200$ (Rational Survivors):} M1 Utility; Reality Parameters ($\alpha=0.665, \sigma=4.9$).
    \item \textbf{$B \ge 200$ (Biased Strivers):} M4b Utility (DP); Belief Parameters ($\alpha=0.5, \sigma=0.4$).
\end{itemize}

\subsection{The ``U-Shaped'' Fertility Landscape}

\begin{figure}[htbp]
    \centering
    \includegraphics[width=1.0\textwidth]{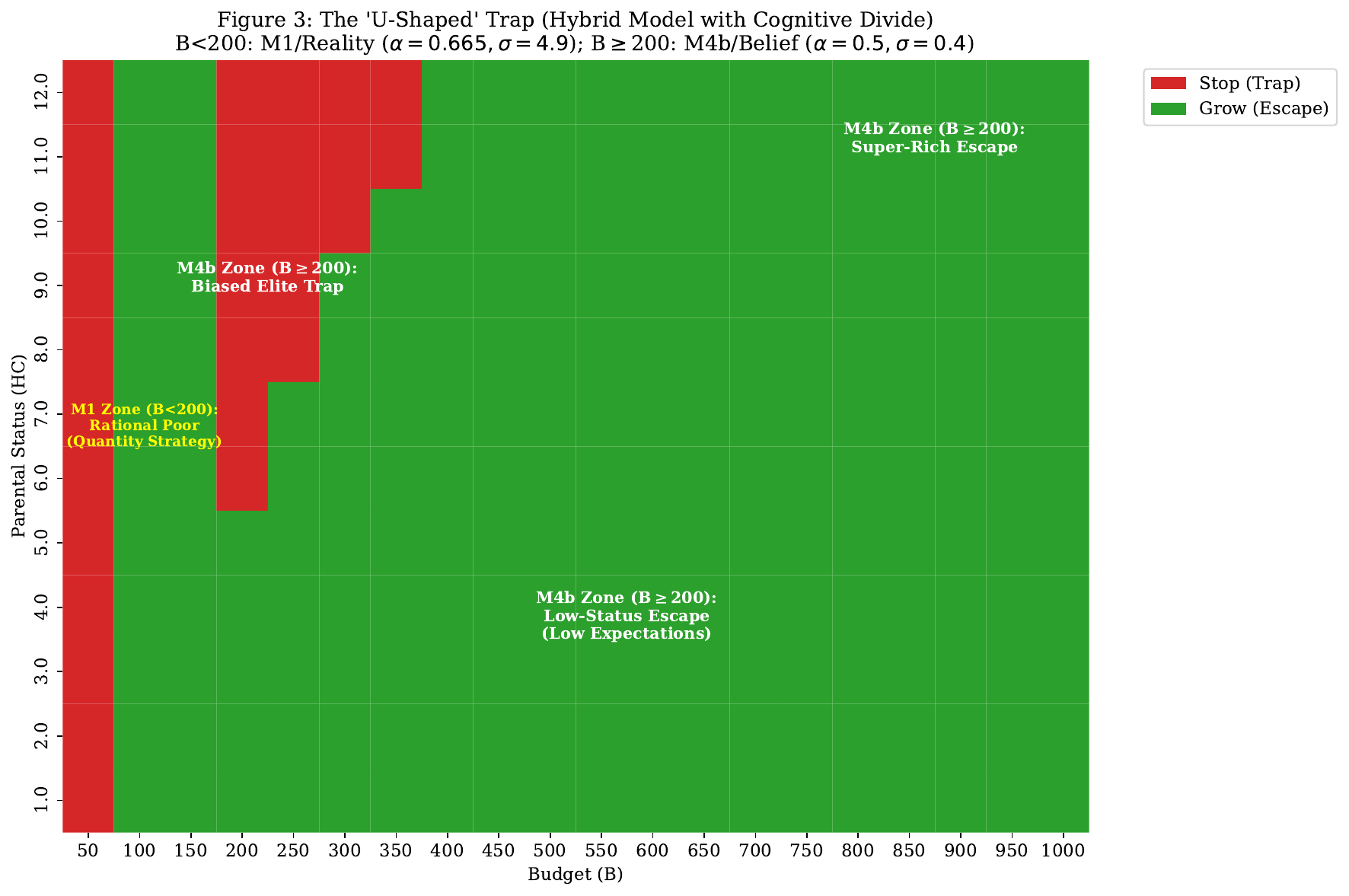}
    \caption{The ``U-Shaped'' Trap (Hybrid Model with Cognitive Divide). $B<200$: M1/Reality; $B \ge 200$: M4b/Belief. The red triangular area represents the ``Competence Trap'' for the aspirational middle class.}
    \label{fig:fig3}
\end{figure}

Figure \ref{fig:fig3} presents the central result: a heatmap illustrating the decision to ``Stop'' (Red) or ``Grow'' (Green).

\textbf{Crucial Clarification: The Activation of the ``Marginal Error''.} The primary driver of the trap for the $B \ge 200$ group is the activation of the Marginal Error. Agents perceive a low return ($\alpha=0.5$), significantly lower than reality ($\alpha=0.665$). This acts as a binding constraint.

\textbf{Zone 1: The Rational Survivors ($B < 200$).} The region below $B=200$ is largely Green (``Grow''), except at the lowest budget ($B=50$) due to fixed costs. This confirms that the poor escape the behavioral fertility trap by retaining a Rational Survival Strategy (M1), prioritizing diversification against massive real risk.

\textbf{Zone 2: The Biased Strivers ($B \ge 200$).} Upon crossing the threshold, agents enter the ``Status Game.''
\begin{itemize}
    \item \textbf{Low-Status Middle ($HC \le 4$):} Escape via Low Expectations. Their aspirations ($R$) are modest, making the goal achievable even with low perceived returns.
    \item \textbf{Aspirational Middle Class (The Competence Trap):} A distinct triangular Red area emerges. This group embodies the ``Curse of Competence.'' Their high HC generates high aspirations ($R$). When combined with the pessimistic Marginal Error, concentrating resources on a single child becomes the only viable strategy.
    \item \textbf{The Rich Escape ($B \ge 400$):} Sufficient wealth relaxes the budget constraint, allowing elites to overcome the trap.
\end{itemize}

\subsection{The Micro-Logic of the Trap}

\begin{figure}[htbp]
    \centering
    \includegraphics[width=1.0\textwidth]{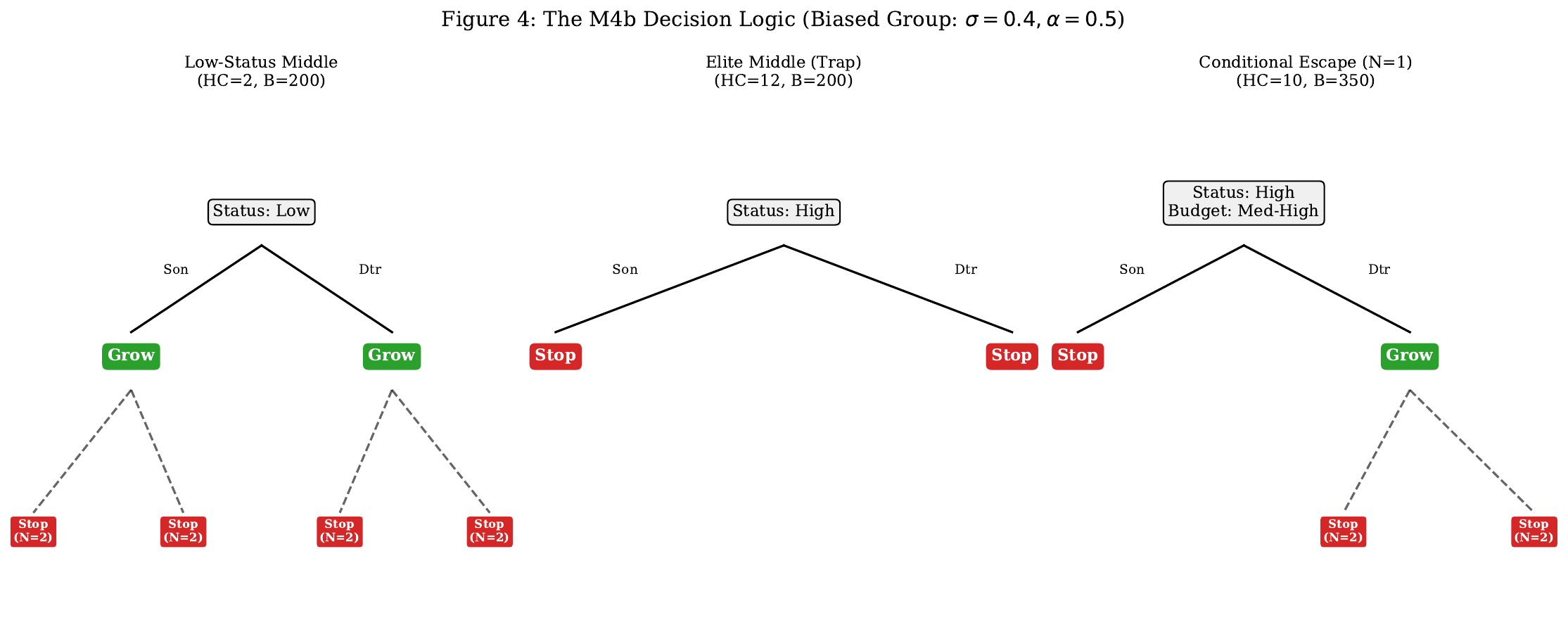}
    \caption{The M4b Decision Logic (Biased Group). It illustrates how the macro patterns emerge from micro-level choices involving gendered aspirations.}
    \label{fig:fig4}
\end{figure}

Figure \ref{fig:fig4} visualizes the sequential decision paths for the Biased Strivers.

\begin{itemize}
    \item \textbf{Case 1: Low-Status Middle ($HC=2, B=200$).} Grow at $N=1$. Low aspirations make the status goal easily achievable.
    \item \textbf{Case 2: Elite Middle Trap ($HC=12, B=200$).} Universal Stop at $N=1$. Aspirations are extremely high. Given the low perceived return ($\alpha=0.5$), $V^{Stop}$ decisively dominates.
    \item \textbf{Case 3: Conditional Escape ($HC=10, B=350$).} The role of gendered aspirations. If the first child is a son (High $R$), parents Stop. If a daughter (Lower $R$), parents Grow.
\end{itemize}

\subsection{Synthesis: The Cognitive Class Divide}

The Hybrid Model resolves the dual puzzles. The ``Ancient vs. Modern'' puzzle is resolved by the Causal Error (biased risk perception). The ``Class Divide'' puzzle is resolved by Cognitive Heterogeneity. The involution trap specifically targets those who enter the status game and adopt its associated cognitive distortions.
\section{Discussion and Robustness}

\subsection{Robustness of the Hybrid Framework}

We validated the model through counterfactual simulations (see Appendix A and B).
\begin{itemize}
    \item \textbf{Necessity (Appendix A.1, A.4):} Uniform cognitive models fail. A pure M4b model predicts a ``Poverty Trap,'' while a pure Reality model predicts ``Universal Stop.''
    \item \textbf{Parameter Stability (Appendix A.2, A.3, A.5):} The trap is robust to variations in belief parameters. Even if the Causal Error is weakened ($\sigma_{Belief}=1.0$), the trap persists due to the Marginal Error.
    \item \textbf{Institutional Models (Appendix B):} The alternative hypothesis of objective diminishing returns (M4c) fails to replicate the U-shape. It either creates a Permanent Elite Trap (High Penalty) or Universal Growth (Low Penalty).
\end{itemize}

\subsection{Testable Predictions}

The model generates distinct predictions:
\begin{enumerate}
    \item \textbf{Cognitive Discontinuity:} There should be a sharp discontinuity in risk and return perception around the status-game entry threshold.
    \item \textbf{Marginal Error and Competition:} Perceived returns ($\alpha_{Belief}$) should be negatively correlated with local competition intensity.
    \item \textbf{Competence Trap:} Controlling for budget, higher Human Capital among the middle class should be associated with \textit{lower} fertility.
\end{enumerate}

\subsection{Policy Implications}

Traditional subsidies may be ineffective if the binding constraint is the pessimistic belief about returns. Effective policies must target cognitive biases:
\begin{itemize}
    \item \textbf{Combating the Marginal Error:} Educational reforms to reduce zero-sum competition and increase \textit{perceived} returns.
    \item \textbf{Combating the Causal Error:} Correcting the illusion of control by highlighting the role of luck, though self-serving biases are hard to de-bias.
\end{itemize}

\subsection{Limitations}

Finally, we acknowledge certain simplifications in our structural assumptions.

First, we model the cognitive boundary as a strict budget threshold ($B \ge 200$) for simplicity. Theoretically, entry into the ``Status Game'' (and the adoption of M4b preferences) might also depend on cultural capital or parental human capital. However, our simulations show that for low-HC families who cross the budget threshold, the low aspirational reference point ($R$) inherently generates a ``Grow'' strategy within the M4b framework, behaviorally converging with the M1 outcome. Thus, making the boundary endogenous to HC would add computational complexity without altering the qualitative U-shaped prediction.

Second, we assume uniform objective risk ($\sigma_{Real}$) across classes to isolate the role of cognitive heterogeneity. Future research could structurally estimate whether financial wealth acts as an implicit insurance mechanism that objectively lowers $\sigma$ for the elite. However, provided that stochastic developmental shocks are largely structural \citep{Agostinelli2025}, any risk-reduction from wealth is likely partial. Consequently, the central mechanism of our model—the entrapment of the aspirational middle class due to the collision of high expectations and constrained resources—remains robust.

\section{Conclusion}

This study resolves the puzzle of modern reproductive strategies by proposing a framework rooted in \textbf{Cognitive Heterogeneity}. We argue that the ``Involution Trap'' is a precise behavioral error driven by class-specific biases.

Our \textbf{Hybrid Model} demonstrates that:
\begin{itemize}
    \item The shift from Quantity to Quality is driven by a \textbf{Causal Error} (underestimating risk).
    \item The U-shaped pattern is driven by a \textbf{Cognitive Divide}. The Poor remain rational survivors. The Aspirational Middle Class is trapped by a \textbf{Two-Stage Belief Error}, where high aspirations collide with a pessimistic belief in low marginal returns (\textbf{The Competence Trap}).
\end{itemize}

Modern involution is not an inevitable economic outcome but a behavioral ``Curse of Competence.'' Trapped in a belief world where control is overestimated and returns are underestimated, strivers rationally optimize within an irrational framework.

\section*{Acknowledgements}
The core idea, hypotheses, and real-world observations originated entirely from the author. The author acknowledges the assistance of Google's Gemini in code debugging and language editing. The author bears full responsibility for all viewpoints and errors.

\bibliographystyle{plainnat}
\bibliography{references}

\appendix
\numberwithin{figure}{section}

\section{Appendix A: Necessity and Robustness of the Hybrid Model}

This appendix provides a series of counterfactual simulations and sensitivity analyses to validate the core mechanisms of the Hybrid Model (M-H) presented in the main text. We test the necessity of cognitive heterogeneity, the robustness of the behavioral parameters, and the stability of the model structure.

\subsection{Necessity Check: The Pure M4b Counterfactual}

\textbf{Purpose:} To demonstrate that a uniform cognitive model, where all agents adopt the M4b (Anxiety) framework and the ``Belief'' parameters ($\alpha=0.5, \sigma=0.4$), cannot explain the observed U-shaped fertility pattern.

\textbf{Setup:} We simulate the entire population (all $B$, all $HC$) using the M4b Dynamic Programming solver with the ``Belief'' parameters.

\begin{figure}[htbp]
    \centering
    \includegraphics[width=0.95\textwidth]{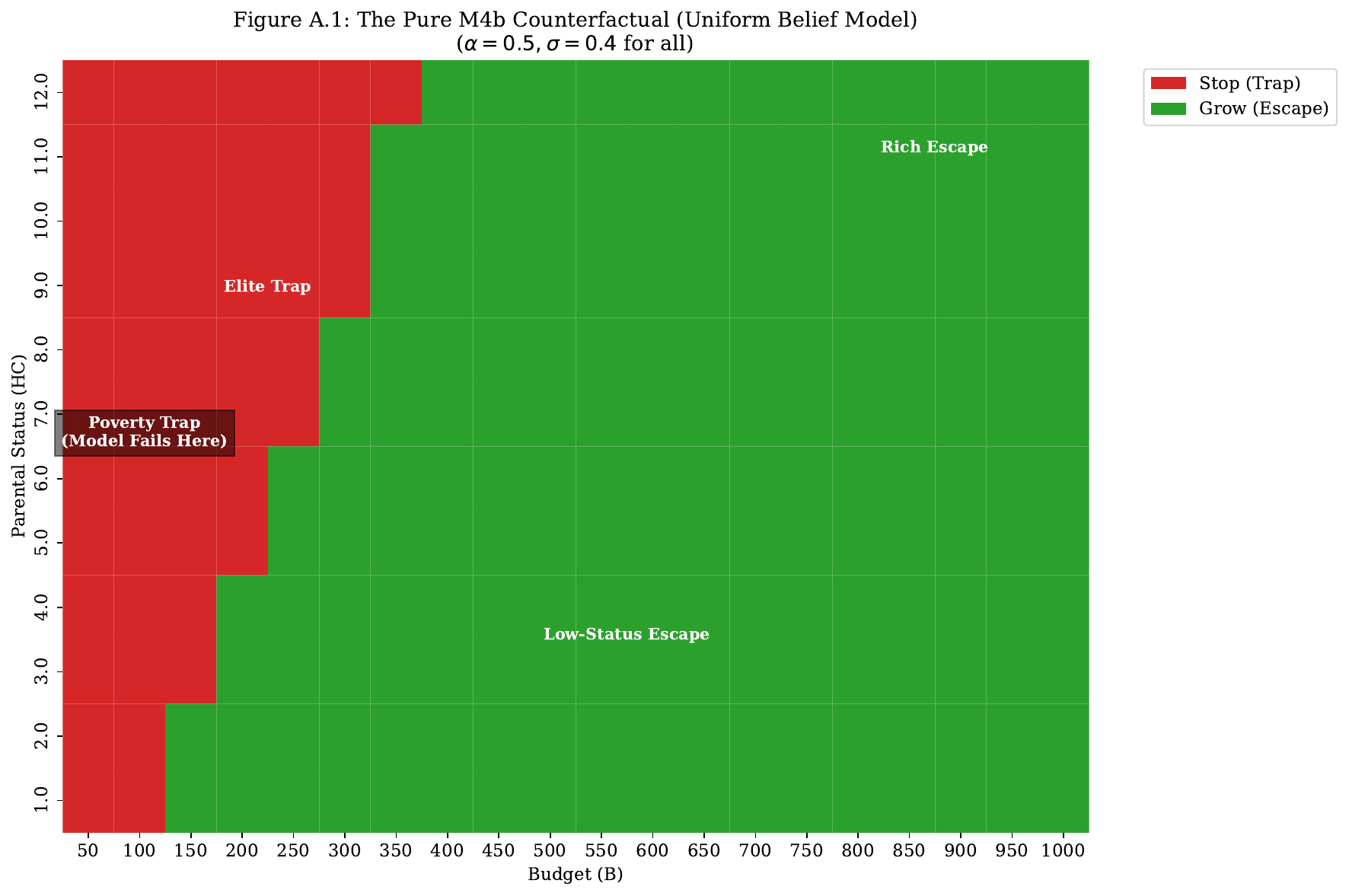}
    \caption{The Pure M4b Counterfactual (Uniform Belief Model). The model predicts a ``Poverty Trap'' (Red area at $B<200$), contradicting the observed resilience of the poor.}
    \label{fig:A1}
\end{figure}

\textbf{Results (Figure \ref{fig:A1}):} The simulation results deviate significantly from the Hybrid Model in the low-budget region ($B<200$). When the poor utilize the M4b framework, they fall into a ``Poverty Trap.'' Under M4b, the combination of low budget and status anxiety (loss aversion) makes having more than one child prohibitively expensive. This contradicts empirical observations, confirming that the M1/Reality regime is necessary to explain the behavior of the poor.

\subsection{Parameter Robustness: 2D Sensitivity Scan}

\textbf{Purpose:} To test the robustness of the ``Competence Trap'' mechanism within the Biased Strivers group ($B \ge 200$) to variations in the belief parameters ($\alpha_{Belief}, \sigma_{Belief}$).

\textbf{Setup:} We perform a 2D sensitivity scan for a representative middle-class agent ($HC=6, B=200$), varying $\alpha$ from 0.3 to 0.8 and $\sigma$ from 0.1 to 1.5.

\begin{figure}[htbp]
    \centering
    \includegraphics[width=0.85\textwidth]{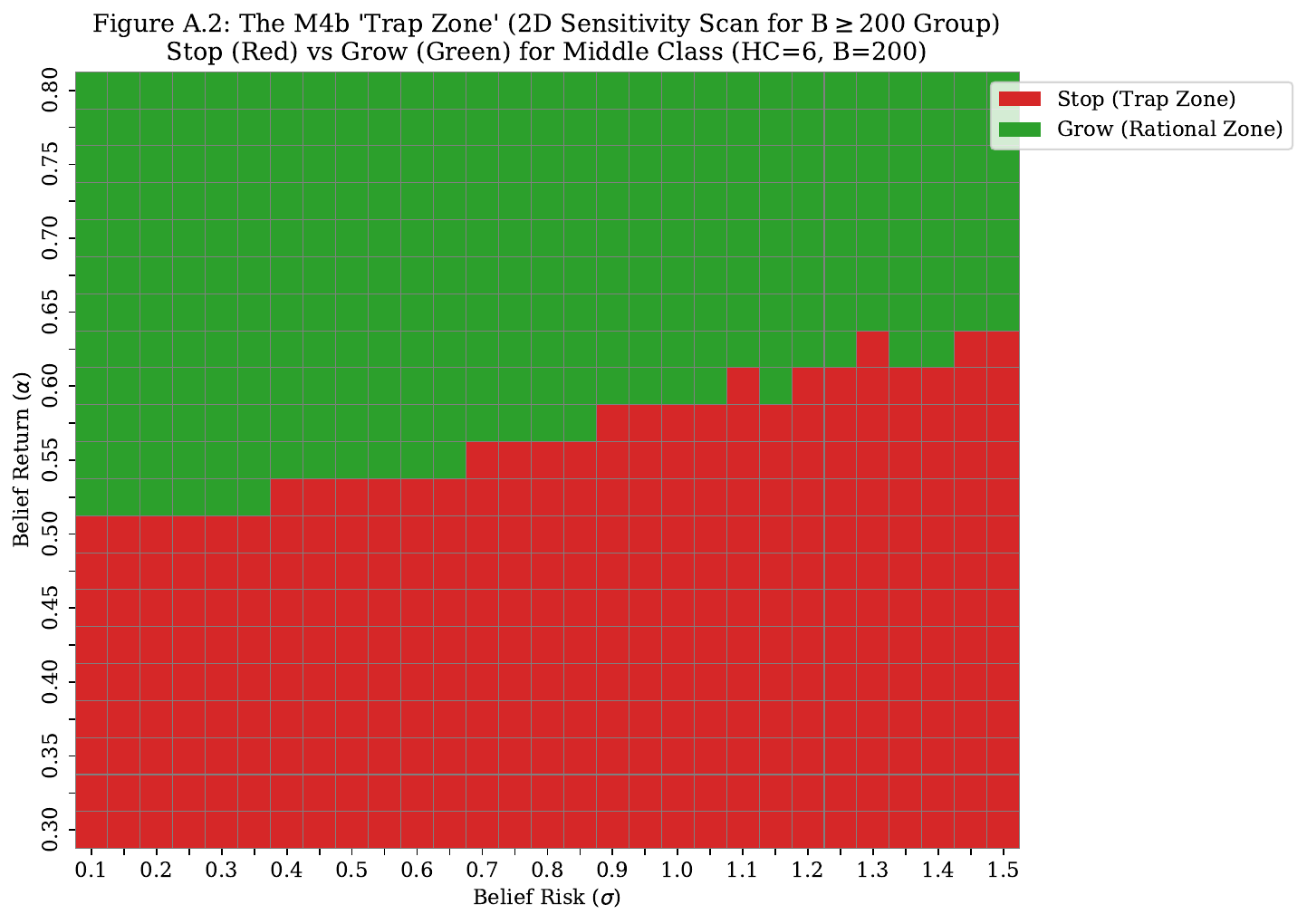}
    \caption{The M4b ``Trap Zone'' (2D Sensitivity Scan for $B \ge 200$, $HC=6$). The trap is stable across a wide range of belief parameters.}
    \label{fig:A2}
\end{figure}

\textbf{Results (Figure \ref{fig:A2}):} The heatmap reveals a large, stable ``Trap Zone'' (Red area). The trap is highly sensitive to the Marginal Error (low $\alpha$). When $\alpha$ is low, the trap persists even if perceived risk ($\sigma$) is high. This indicates that the involution mechanism is not dependent on the precise calibration of these parameters but represents a stable behavioral phenomenon.

\subsection{Hybrid Robustness: High Belief Risk}

\textbf{Purpose:} To test the stability of the Hybrid Model if the ``Causal Error'' is weakened—i.e., if the Biased Strivers perceive a higher level of risk.

\textbf{Setup:} We maintain the Hybrid Model structure but increase the perceived risk for Strivers ($B \ge 200$) significantly from $\sigma_{Belief}=0.4$ to $\sigma=1.0$, while keeping $\alpha_{Belief}=0.5$.

\begin{figure}[htbp]
    \centering
    \includegraphics[width=0.95\textwidth]{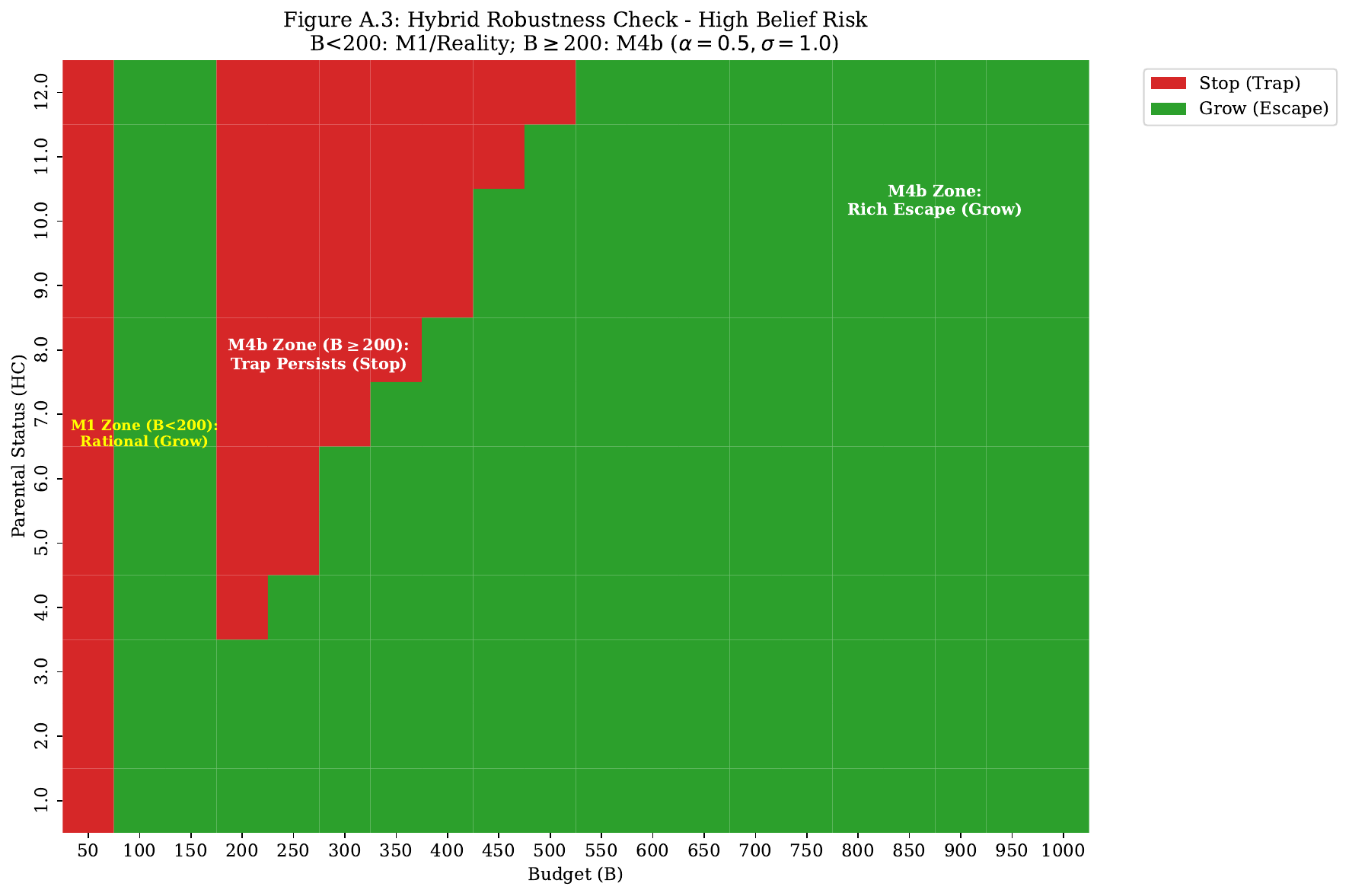}
    \caption{Hybrid Robustness Check - High Belief Risk ($\sigma=1.0$). The trap persists due to the binding constraint of the Marginal Error ($\alpha=0.5$).}
    \label{fig:A3}
\end{figure}

\textbf{Results (Figure \ref{fig:A3}):} Despite the significant increase in perceived risk, the ``Competence Trap'' remains largely intact. This highlights the dominant role of the ``Marginal Error'' ($\alpha=0.5$). When perceived returns are low, the constraint on fertility remains binding, even if the illusion of control is weakened.

\subsection{Necessity Check: The ``Real World'' Counterfactual}

\textbf{Purpose:} To demonstrate that the difference in behavior between the Poor and the Middle Class requires a difference in utility frameworks (M1 vs. M4b), not just parameter perceptions.

\textbf{Setup:} We simulate the population using M4b preferences but assuming everyone correctly perceives Reality parameters ($\alpha=0.665, \sigma=4.9$).

\begin{figure}[htbp]
    \centering
    \includegraphics[width=0.95\textwidth]{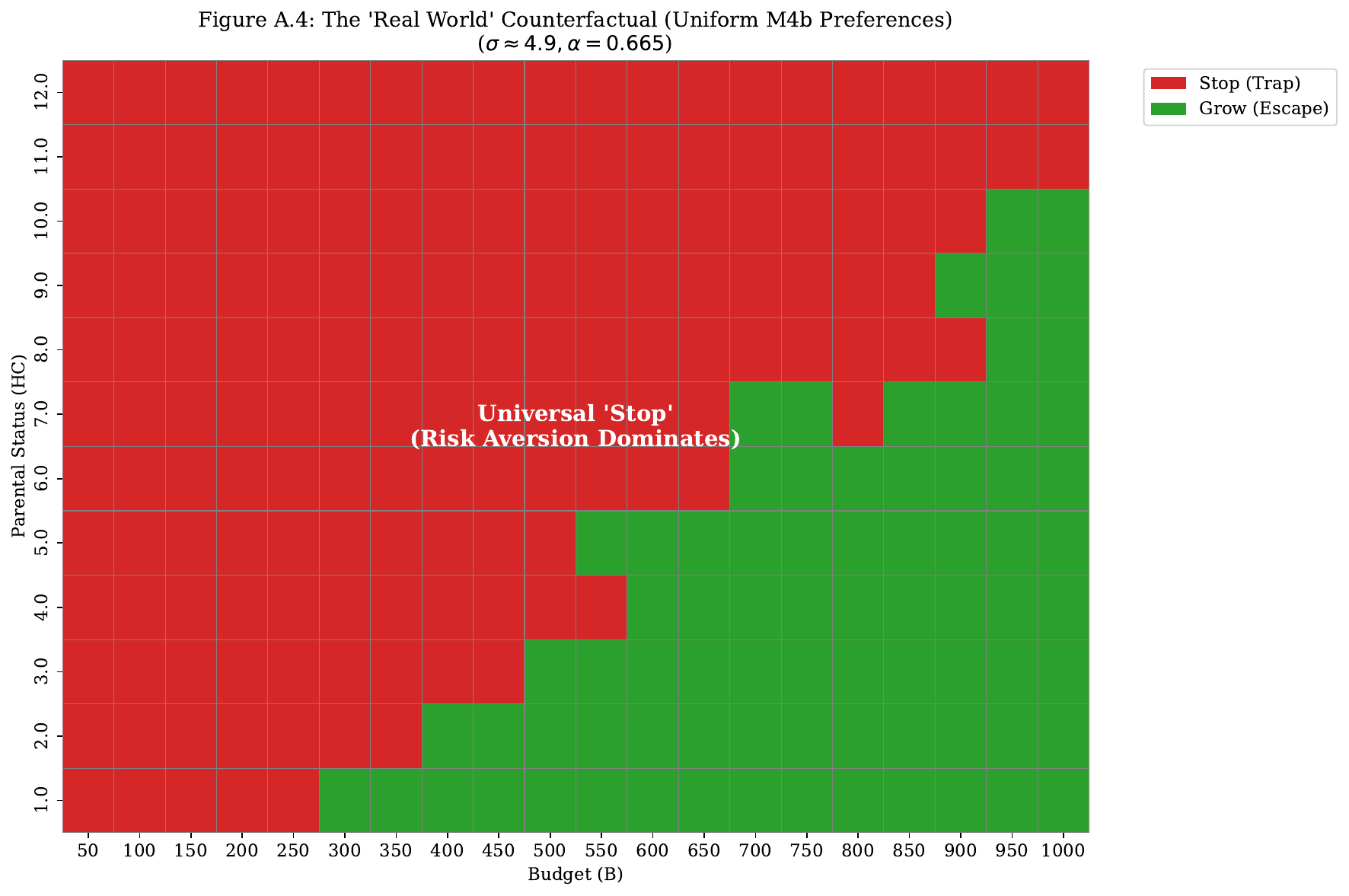}
    \caption{The ``Real World'' Counterfactual (Uniform M4b, $\sigma \approx 4.9$). The result is a Universal Stop, failing to explain the behavior of the poor.}
    \label{fig:A4}
\end{figure}

\textbf{Results (Figure \ref{fig:A4}):} The simulation shows a ``Universal Stop.'' The extreme risk aversion inherent in M4b, when exposed to massive real-world risk, paralyzes decision-making. This proves that the Poor's escape requires the adoption of the M1 (Survival) utility framework.

\subsection{Structural Robustness: Shifting the Boundary}

\textbf{Purpose:} To verify that the qualitative results are not dependent on the precise location of the Cognitive Boundary ($B=200$).

\textbf{Setup:} We shift the Cognitive Boundary from $B=200$ to $B=250$.

\begin{figure}[htbp]
    \centering
    \includegraphics[width=0.95\textwidth]{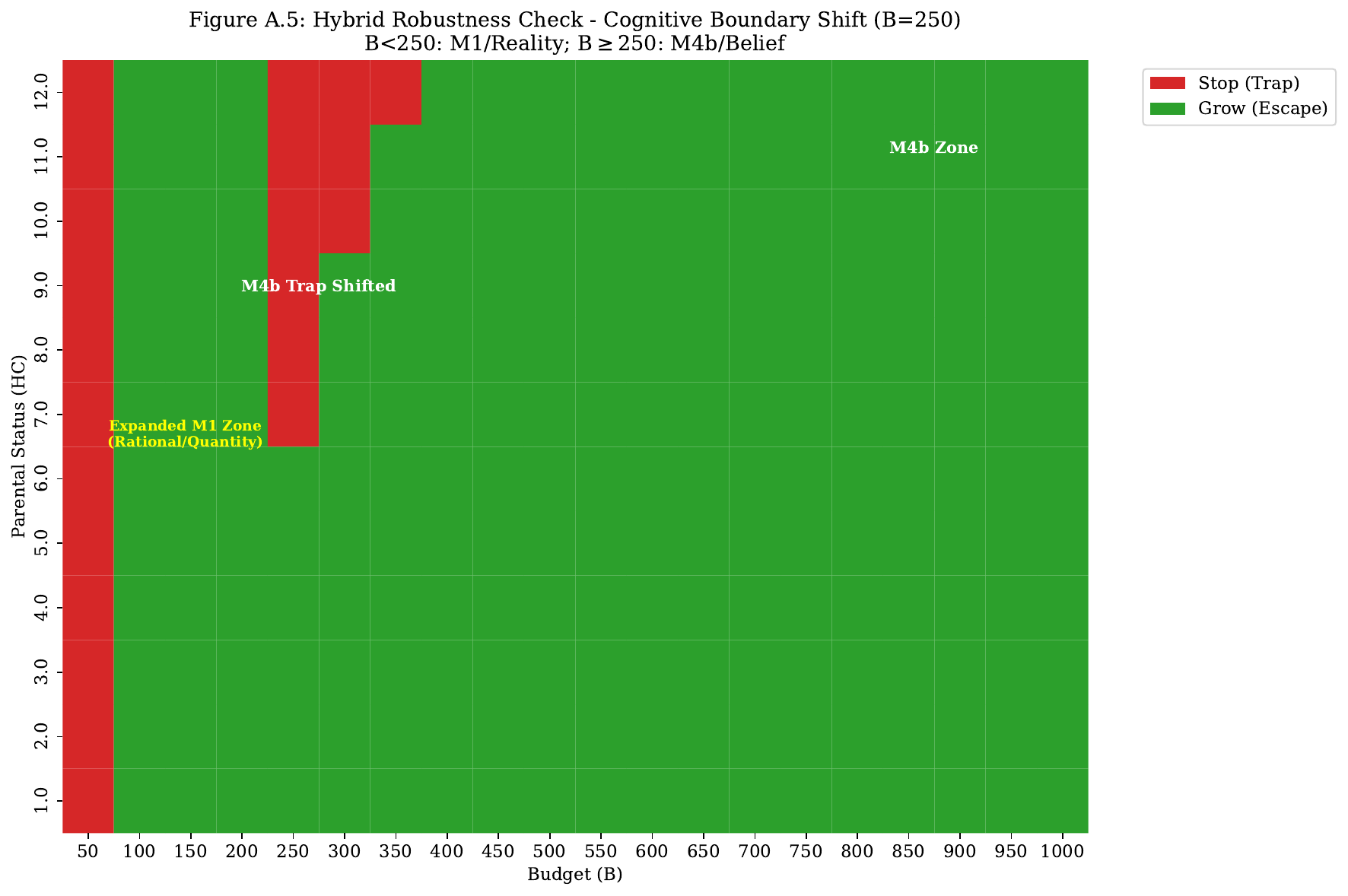}
    \caption{Hybrid Robustness Check - Cognitive Boundary Shift ($B=250$). The U-shaped pattern is structurally robust.}
    \label{fig:A5}
\end{figure}

\textbf{Results (Figure \ref{fig:A5}):} The overall U-shaped pattern persists, merely shifting rightward. The existence of a cognitive divide, rather than its exact location, drives the results.

\section{Appendix B: Alternative Explanations – The Hybrid Institutional Model (M1/M4c)}

This appendix tests the competing hypothesis that the trap is driven by objective diminishing returns (Institutional Model M4c) rather than behavioral biases.

\subsection{Model Setup: The M4c Framework}
To test this hypothesis, we introduce Model 4c (M4c), the ``Institutional'' framework. M4c replaces the linear production function of M4b with a quadratic function (in log space), which explicitly models diminishing returns:
\begin{equation}
    \ln(HC_i) = \ln(A) + \alpha_1 \ln(I_i) - \alpha_2 (\ln(I_i))^2 + \ln(\varepsilon_i)
\end{equation}
The parameter $\alpha_2 > 0$ represents the degree of diminishing returns (the ``penalty'').

\subsection{Scenario 1: High Penalty (\texorpdfstring{$\alpha_2=0.05$}{alpha2=0.05})}

\begin{figure}[htbp]
    \centering
    \includegraphics[width=0.95\textwidth]{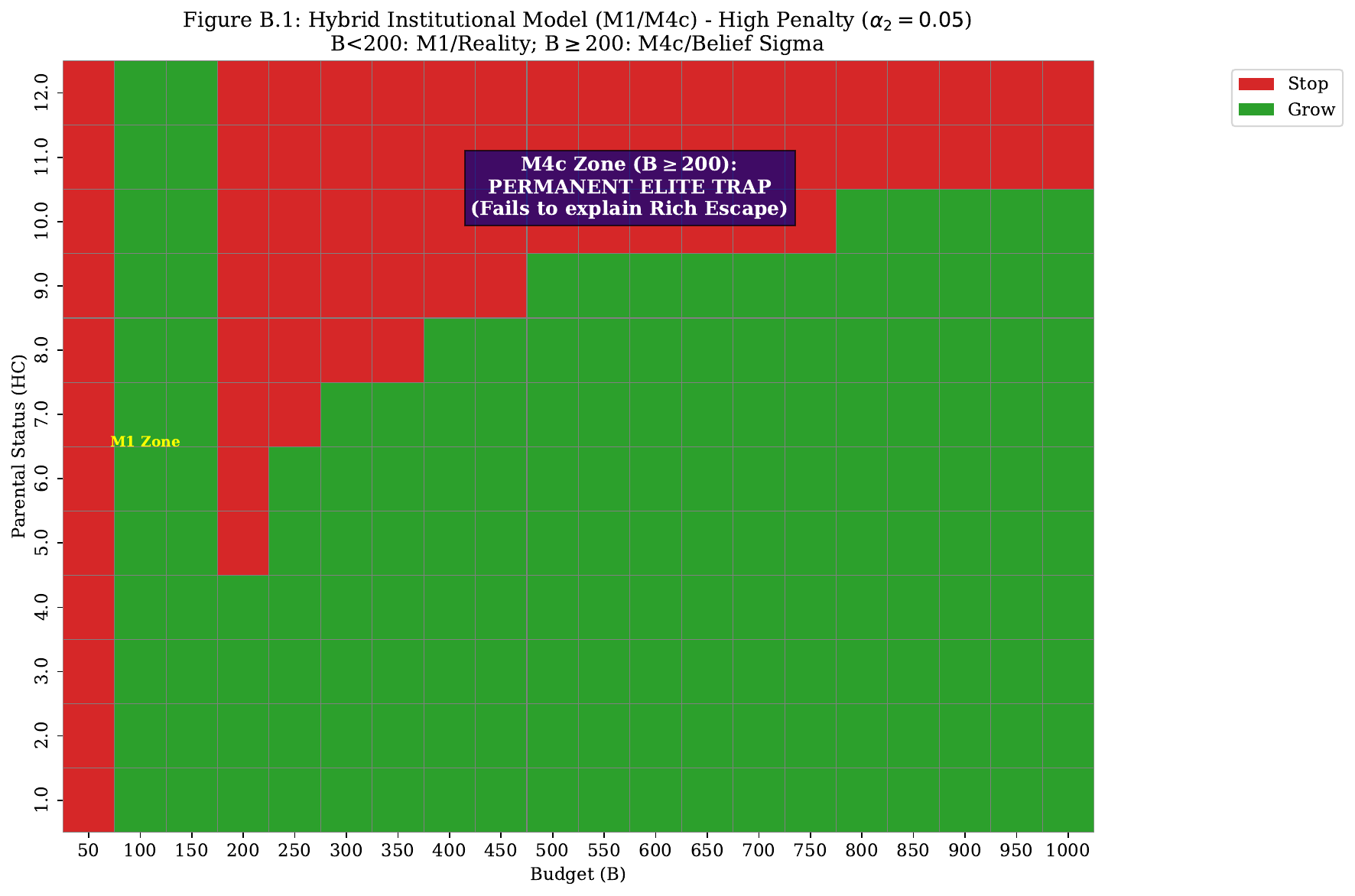}
    \caption{Hybrid Institutional Model - High Penalty. It creates a ``Permanent Elite Trap,'' failing to explain the escape of the rich.}
    \label{fig:B1}
\end{figure}

\textbf{Results (Figure \ref{fig:B1}):} The simulation shows a ``Permanent Elite Trap.'' When $\alpha_2$ is high, it becomes virtually impossible for high-HC parents to achieve their aspirations, regardless of budget. This contradicts reality as it fails to explain the rich escape.

\subsection{Scenario 2: Low Penalty (\texorpdfstring{$\alpha_2=0.02$}{alpha2=0.02})}

\begin{figure}[htbp]
    \centering
    \includegraphics[width=0.95\textwidth]{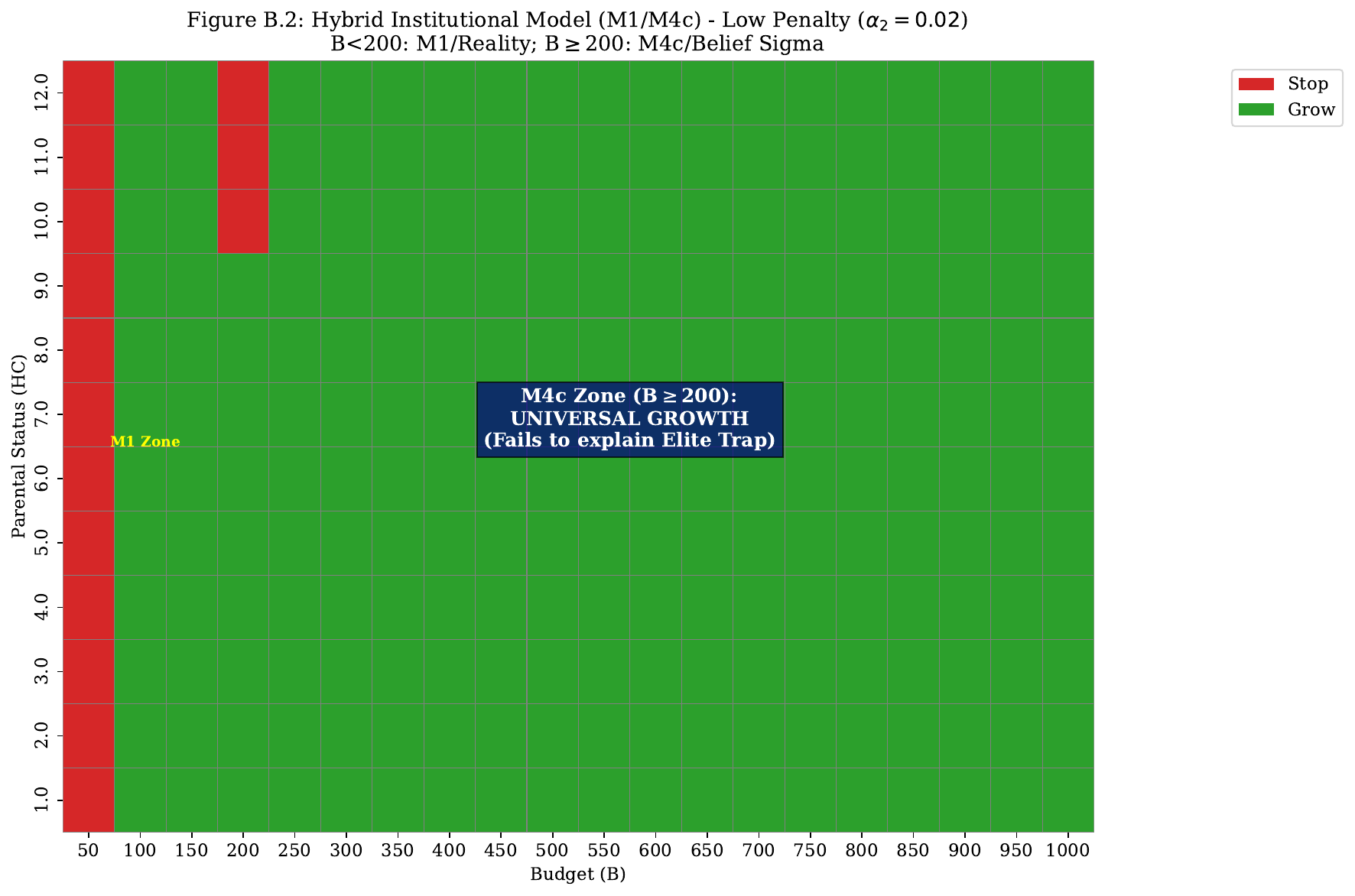}
    \caption{Hybrid Institutional Model - Low Penalty. It results in ``Universal Growth,'' failing to explain the existence of the trap.}
    \label{fig:B2}
\end{figure}

\textbf{Results (Figure \ref{fig:B2}):} The simulation shows ``Universal Growth.'' When diminishing returns are mild, the objective returns are sufficiently high to allow nearly all families to achieve aspirations. This fails to explain the elite trap.

\subsection{Conclusion on Institutional Models}

The Institutional Framework operates as a blunt instrument. It lacks the precision to specifically target the ``Aspirational Middle Class.'' Furthermore, \citet{Agostinelli2025} find increasing returns to scale at early ages, structurally rejecting the diminishing returns hypothesis. Since the objective technology is characterized by high returns and high volatility, the observed fertility trap must be driven by behavioral mechanisms modeled in our M4b framework.

\section{Appendix C: Methodological Note on the Calibration of Stochastic Risk}

A core component of our ``Belief vs. Reality'' framework is the comparison between the parents' perceived low risk ($\sigma_{Belief} \approx 0.4$) and the ``massive'' empirical risk ($\sigma_{Real}$) identified by \citet{Agostinelli2025}.

A reviewer might correctly note that the \citet{Agostinelli2025} model is a high-dimensional structural model, and its specific variance estimates (e.g., the $\sigma=4.922$ we use in our own ``Real World'' simulation in Appendix A.4) are not directly transferable (i.e., ``plug-and-play'') into our more stylized theoretical framework. The $\sigma$ parameter in our model represents a single, unified source of stochastic shock, whereas in \citet{Agostinelli2025} it represents a complex, multi-factor variance structure.

We must clarify that our primary use of the \citet{Agostinelli2025} $\sigma$ estimate is \textit{qualitative}, not quantitative, and serves a specific, two-step purpose in our argument:

\begin{enumerate}
    \item \textbf{Step 1 (Internal Model Threshold):} First, our own static simulations (Section 4, Figure 1) establish a critical internal threshold. We demonstrate that the M2 ``Anxiety'' (Quality) strategy is rationally viable only in a perceived low-risk world. Specifically, for a representative middle-class household ($HC=6$), the rational switching threshold is $\sigma^* \approx 0.45$. As soon as uncertainty exceeds this level, the M1 ``Survival'' (Quantity) strategy rationally dominates.
    
    \item \textbf{Step 2 (External Benchmark Validation):} Second, the \citet{Agostinelli2025} paper serves as our external benchmark to determine which regime (``low-risk'' or ``high-risk'') constitutes ``Empirical Reality.'' Their finding that stochastic shocks are ``massive'' (e.g., their estimate of $\sigma \approx 4.9$ is an order of magnitude larger than our model's $0.45$ threshold) provides definitive evidence that the ``Real World'' is objectively a high-uncertainty environment.
\end{enumerate}

Therefore, the \citet{Agostinelli2025} finding is not used to quantitatively calibrate our M1 model's precise utility function. Rather, it is used as the external justification to qualitatively validate the M1 (Quantity) model's relevance as the objective rational baseline. This two-step process establishes our central puzzle: If ``Reality'' is a high-$\sigma$ world (per A\&W) and the rational response in such a world is the M1 (Quantity) strategy (per our Figure 1), then why do we observe modern elites universally adopting the M2 (Quality) strategy? This gap is precisely what necessitates our ``Two-Stage Belief Error'' explanation.
\end{document}